\documentclass[%
aip,
rsi,
 amsmath,amssymb,
 reprint,%
]{revtex4-2}
\usepackage{xcolor}
\usepackage{soul}
\usepackage{multirow}
\colorlet{soulyellow}{yellow!0} 
\sethlcolor{soulyellow}%
\usepackage{graphicx}
\usepackage{float}
\usepackage{bm}
\usepackage{mathrsfs}

\usepackage[utf8]{inputenc}
\usepackage[T1]{fontenc}
\usepackage{mathptmx}
\usepackage{etoolbox}

\renewcommand{\selectlanguage}[1]{}

\usepackage{hyperref}

\makeatletter
\def\@email#1#2{%
 \endgroup
 \patchcmd{\titleblock@produce}
  {\frontmatter@RRAPformat}
  {\frontmatter@RRAPformat{\produce@RRAP{*#1\href{mailto:#2}{#2}}}\frontmatter@RRAPformat}
  {}{}
}%

\makeatother
\begin{document}

\preprint{AIP/123-QED}

\title{Stabilizing an optical cavity containing a bulk diamond crystal at millikelvin temperatures in a cryogen-free dilution refrigerator}

\author{Tatsuki Hamamoto}
\thanks{These two authors contributed equally.}
 \affiliation{Experimental Quantum Information Physics Unit, Okinawa Institute of Science and Technology Graduate University, Onna, Okinawa 904-0495, Japan}

\author{Amit Bhunia}%
\thanks{These two authors contributed equally.}
 \affiliation{The Science and Technology Group, Okinawa Institute of Science and Technology Graduate University, Onna, Okinawa 904-0495, Japan}

\author{Hiroki Takahashi}%
 \affiliation{Experimental Quantum Information Physics Unit, Okinawa Institute of Science and Technology Graduate University, Onna, Okinawa 904-0495, Japan}

\author{Yuimaru Kubo}
 \affiliation{The Science and Technology Group, Okinawa Institute of Science and Technology Graduate University, Onna, Okinawa 904-0495, Japan}
 \email{yuimaru.kubo@oist.jp}
\date{\today}

\begin{abstract}
We successfully stabilized a Fabry-P\'erot optical cavity incorporating a bulk diamond crystal at millikelvin temperatures in a cryogen-free dilution refrigerator with the pulse-tube cryocooler running. 
In stark contrast to previous demonstrations where lasers were locked to the cavities, our setup locks the cavity to the laser, thereby ensuring that emitted photons remain at a stable, predetermined frequency. This stability is advantageous for efficient quantum networking applications (e.g. remote entanglement generation) that require indistinguishable photons.
Our measurements of cavity length fluctuation suggest that the setup could stabilize a cavity up to a finesse of $1.3\times 10^4$ without the diamond and $5.3 \times10^3$ with the diamond crystal. 
The finesse with a diamond crystal of approximately 90 is primarily limited by the absorption loss inside the diamond. 
\end{abstract}

\maketitle

\section{\label{sec:introduction}INTRODUCTION}
Optical cavities have diverse applications across various fields. 
They are critical in enhancing the stability and precision of lasers, \cite{boyd_basic_2024} which is vital in optical communications. 
In metrology, they are used for highly accurate measurements, essential in both scientific research\cite{aspelmeyer_cavity_2014} and industrial applications. 
In quantum technology, optical cavities facilitate the control and manipulation of quantum states in the framework of cavity quantum electrodynamics (cavity-QED). \cite{mabuchi_cavity_2002,walther_cavity_2006} 
Additionally, they will play an essential role in quantum network nodes that achieve quantum communication and/or distributed quantum computing. \cite {reiserer_cavity-based_2015,reiserer_colloquium_2022} 
Apart from traditional applications, optical cavities operating at cryogenic temperatures, typically around 4 K, are also essential for a variety of quantum technologies such as a single photon source, \cite{lodahl_interfacing_2015,senellart_high-performance_2017} ion trap, \cite{wipfli_integration_2023} optomechanics,\cite{zhong_millikelvin_2017, ruelle_tunable_2022} microwave-millimeter wave converter \cite{kumar_quantum-enabled_2023} or a gravitational wave detector.\cite{akutsu_large-scale_2015}

Traditionally, the optical cavities in such high-demand applications have been realized on stabilized optical tables to ensure the necessary mechanical stability and isolation from environmental vibrations at room temperature. 
Besides, the last two decades have witnessed significant advancements in microwave quantum devices operating at millikelvin temperatures in dilution refrigerators, paving the way for microwave-optical hybrid applications, such as quantum transducers.\cite{stannigel_optomechanical_2010,rabl_quantum_2010,tsang_cavity_2010,safavi-naeini_proposal_2011,wang_using_2012,hisatomi_bidirectional_2016,javerzac-galy_-chip_2016,zhong_proposal_2020,williamson_magneto-optic_2014,barnett_theory_2020} 
These devices bidirectionally convert optical and microwave photons, which is crucial for reliable quantum networks. \cite{wehner_quantum_2018,reiserer_colloquium_2022,ruf_quantum_2021,awschalom_development_2021} 
Locking an optical cavity resonance frequency to the laser frequency is vital for photon indistinguishability in these applications. 
However, stabilizing an optical cavity length within the commonly used `cryogen-free' dilution refrigerators poses significant challenges due to considerable mechanical vibrations from the pulse-tube cryocoolers. \cite{olivieri_vibrations_2017,schmoranzer_cryogenic_2019,chijioke_vibration_2010} 

One approach to mitigate cryocooler vibrations is to integrate optical cavities into on-chip devices (e.g., photonic crystal or whispering-gallery cavities)~\mbox{\cite{mohammad_mirhosseini_superconducting_2020, forsch_microwave--optics_2020, xu_bidirectional_2021,hease_bidirectional_2020,weaver_integrated_2024,rochman_microwave--optical_2023}
} and deliver light via fibers to the mixing chamber. This configuration ensures common-mode motion and suppresses vibration noise, but sacrifices fast tunability of the cavity resonance, necessitating that the laser be locked to the cavity. It also requires precise \textit{in-situ} alignment between the fiber and cavity modes, which can be sensitive to thermal drifts.

In a transducer setup, locking the laser to the cavity causes the frequency of the converted photons to fluctuate over time. Such fluctuations can degrade the fidelity or reduce the generation rate of remote entanglement, required for future quantum network applications. To overcome this, locking the optical cavity to a frequency-stabilized laser is essential for maintaining a stable photon frequency.

Apart from the challenges of stabilizing optical cavities in dilution refrigerators, impurity spins in diamond crystals have emerged as a promising resource for various quantum technology applications, including quantum sensing, \cite{fu_sensitive_2020,aslam_quantum_2023} microwave quantum memory, \cite{kubo_hybrid_2011,zhu_coherent_2011,grezes_multimode_2014} and stable photon sources. \cite{sipahigil_integrated_2016,iwasaki_germanium-vacancy_2015,iwasaki_tin-vacancy_2017} 
These spins are also potential candidates for spin-based microwave-optical photon transducers, which must be placed inside a dilution refrigerator to avoid thermal noise at microwave frequencies. 
In this spin-based approach,\mbox{\cite{williamson_magneto-optic_2014,barnett_theory_2020,fernandez-gonzalvo_cavity-enhanced_2019,king_probing_2021,blum_interfacing_2015,li_quantum_2017,obrien_interfacing_2014}} efficient microwave–optical quantum transduction requires embedding a bulk diamond crystal within the mode of an optical cavity to enable strong spin–photon coupling. 

In this work, we mitigate the vibration issue by ensuring that the optical cavity, mounted at the coldest temperature stage, and the incoming optical beams share common mechanical modes, achieved through an optical breadboard installed on top. 
Using this setup, we reproducibly locked an \textit{optical cavity to the laser} at approximately $15\,\mathrm{mK}$. 
As a first step toward efficient microwave-optical photon quantum transduction\cite{williamson_magneto-optic_2014,barnett_theory_2020,rochman_microwave--optical_2023,fernandez-gonzalvo_cavity-enhanced_2019} using an ensemble of impurity spins in diamond crystals, we incorporated a bulk diamond crystal into the optical cavity and successfully locked it. 
We achieved a root mean square (rms) cavity length fluctuation of approximately tens of picometers during locking. 
This measured cavity length fluctuation indicates that the cavity can readily support an effective finesse of $1.3\times10^4$ (without diamond) and $5.3 \times10^3$ (with a diamond crystal). 
This is the first demonstration of cavity--to--laser locking at millikelvin temperatures in a system integrating a bulk diamond crystal. 
These results pave the way for the development of a diamond-based microwave–optical quantum transducer. 

This paper is organized as follows:
In Section \ref{sec:mechanical_vibration}, we discuss the vibrations of our custom dilution refrigerator setup. 
Absolute and relative vibrations are measured and presented in Subsections \ref{subsec:AbsoluteVibration} and \ref{subsec:RelativeVibration}, respectively. 
In Section \ref{sec:Results_locking}, we describe the optical cavity device and measurement setup in Subsection \ref{subsec:device} and present the results of cavity stabilization, both with and without a diamond crystal, detailed in Subsections \ref{subsec:bareCavLocking} and \ref{subsec:DiaCavLock}, respectively. 
We also investigate the loss mechanism of diamond-integrated cavity in Subsection \ref{subsec:Loss}, the temperature-dependent effects in Subsection \ref{subsec:Cavity_Temperature} and analyze the vibration spectra in Subsection \ref{subsec:VibrationalModes}.

\section{\label{sec:mechanical_vibration}Mechanical Vibration of a Cryogen-Free Dilution Refrigerator}

\subsection{\label{subsec:AbsoluteVibration}Absolute Vibration Measurement}

\begin{figure}[htbp]
    \includegraphics[width=\hsize]{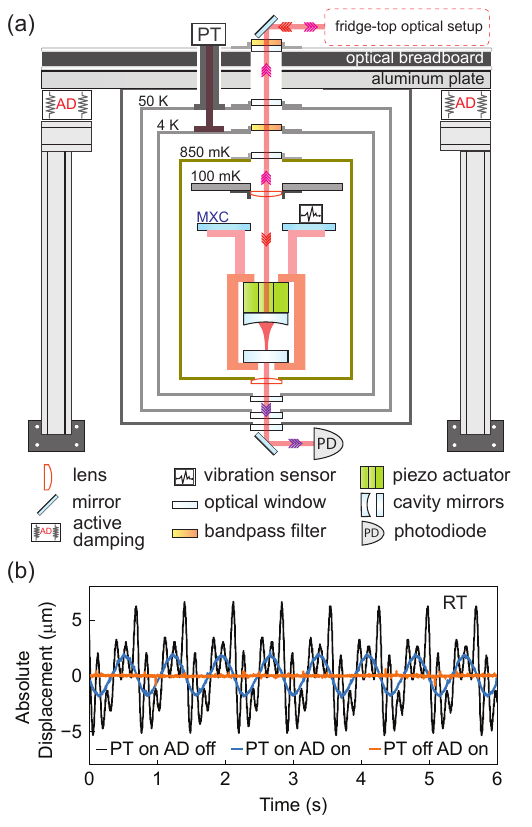}
    \caption{The custom dilution refrigerator setup and vibration measurement. 
    (a) Schematic of the custom dilution refrigerator. 
    An active damping (AD) system supports an aluminum plate, on which both an optical breadboard and the main body of the refrigerator are mounted, suppressing the common-mode vibrations. 
    A laser beam is directed inside via  a 45-degree mirror. 
    Different colored arrows indicate the propagation directions of the input, reflected, and transmitted light. 
    The cavity is thermalized to the mixing chamber (MXC) plate through a cylindrical copper shroud. 
    (b) Absolute vibrations measured on the MXC plate with a vibration sensor under three conditions: PT on and AD off, PT on and AD on, PT off and AD on. 
    }
    \label{fig:1_Fridgesetup} 
\end{figure}

First, we characterize the `absolute' vibrations at the mixing chamber (MXC) stage inside a cryogen-free dilution refrigerator (Bluefors, LD-400) with respect to the earth. 
Figure \ref{fig:1_Fridgesetup}(a) shows the schematic setup of the fridge. 
An optical breadboard is mounted on a $70\,\mathrm{mm}$-thick aluminum plate, which also supports the refrigerator's main body.
This plate is equipped with an active damping (AD) system to minimize the absolute vibrations primarily caused by the pulse-tube cryocooler (PT). 
We measured the absolute vibration levels on the MXC stage using a vibration sensor (Table Stable, VA-2C) under three conditions at room temperature: 
1) PT on and AD off, 2) PT on and AD on, and 3) PT off and AD on. 

The results are presented in Fig. \ref{fig:1_Fridgesetup}(b). Absolute vibration levels are  approximately $1.2\,\mathrm{\mu m}$ (AD on) and $ 2.8\,\mathrm{\mu m}$ (AD off) in root mean square (rms).
The values are an order of magnitude improvement compared to the value reported in Ref.~\onlinecite{olivieri_vibrations_2017} (indicated as ``Hex std'') for a standard configuration of cryogen-free dilution refrigerator similar to ours. 
However, this value remains one to two orders of magnitude larger than those achieved in systems equipped with vibration isolation options (indicated as ``Hex UQT'' in Ref.~\onlinecite{olivieri_vibrations_2017}), where the PT cold head and top flange are isolated via an edge-welded bellow, which is not present in our system.

\subsection{\label{subsec:RelativeVibration}Relative Vibration Measurement}

\begin{figure}[ht]
    \includegraphics[width=\hsize]{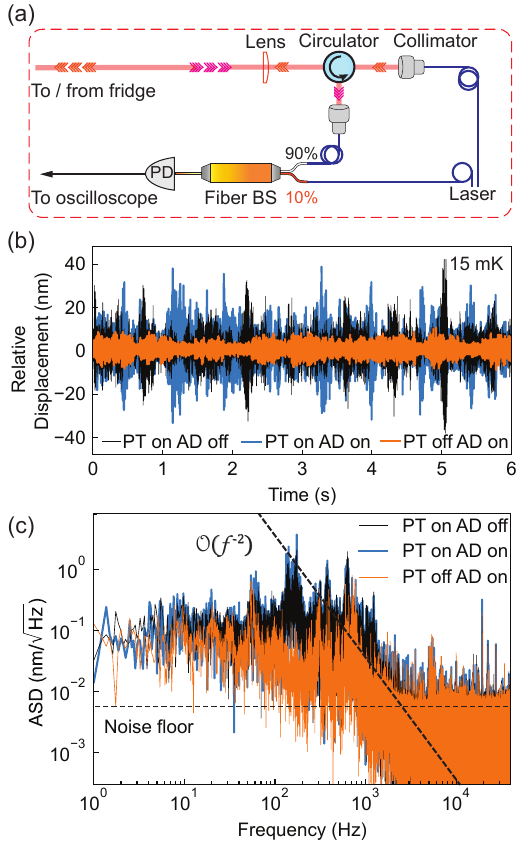}
    \caption{Relative vibration measurement between the device on the MXC plate and the top optical breadboard. 
    (a) Interferometer measurement setup. 
    The reflected laser beam is directed to a fiber beam splitter (BS) through an optical circulator, combined with a local oscillator (LO), and detected by a photodetector (PD). 
    (b) Relative displacement measurements at $15\,\mathrm{mK}$, comparing PT off and AD on (orange), PT on and AD on (blue), PT on and AD off (black). 
    (c) Amplitude spectral density (ASD) derived from the fast Fourier transform (FFT) of the data in (b). 
    The black dashed line is a guide to the eyes for the universal $\mathcal{O} (f^{-2})$ behavior (see main text). 
     }
    \label{fig:2_relative displacement}
\end{figure}

To further mitigate vibrations, optical beam alignment was implemented on the top breadboard. 
This configuration was designed to vibrate the optical cavity on the MXC plate and the fridge's vibrations in a common mode. 
Consequently, only the `relative' vibrations between the fridge-top optical setup and the device at the MXC stage prove to be significant. 
The relative vertical vibration was evaluated by measuring the interference between the reflected laser beam from the optical cavity's input mirror and the local oscillator (LO), as depicted in Fig. \ref{fig:2_relative displacement}(a). 
In this setup, a continuous wave Ti:Sapphire laser (M-squared, SolsTiS) with a wavelength of $737\,\mathrm{nm}$ was used (see Appendix \ref{appendix:subsec:InterferometerSetup} for more details). 
To ensure the laser did not enter the cavity, we introduced a significant misalignment with respect to the cavity mode.

Figure \ref{fig:2_relative displacement}(b) shows the relative vibration measurements at $15\,\mathrm{mK}$ under the conditions PT on and AD off, PT on and AD on, PT off and AD on. 
With PT on, the relative vibration amplitudes are approximately $ 5.6\, \mathrm{nm}$ in rms, which is three orders of magnitude less than the absolute vibration measurements, demonstrating effective vibration mitigation achieved using the optical breadboard on top of the fridge. 
Remarkably, in stark contrast to the absolute vibrations, no significant differences were observed between the active damping (AD) on and off. 
Additionally, the measured vibration amplitudes are nearly identical to those measured at $4\,\mathrm{K}$ without helium mixture circulation (see Fig.\ref{figS:RelVib4K} in Appendix \ref{appendix:subsec:analysis_relVibration}), suggesting the absence of detrimental vibrations from the turbo pump used during dilution refrigerator operations. 

To assess the spectral response of the relative vibrations, we plot the amplitude spectral density (ASD) in Fig. \ref{fig:2_relative displacement}(c), derived from the fast Fourier transform (FFT) of the data in Fig. \ref{fig:2_relative displacement}(b). 
The spectra remain around $\sim 0.1\,\mathrm{nm} \mathrm{Hz}^{-1/2}$ up to approximately $100\,\mathrm{Hz}$, which is four to five orders of magnitude lower than the absolute vibration levels in the low frequency ($\sim \mathrm{Hz}$) region measured on the standard dilution refrigerators (``Hex std'' and ``Triton''in Ref ~\onlinecite{olivieri_vibrations_2017}) and two orders of magnitude lower than even those with vibration isolation option (``Hex UQT'' in Ref ~\onlinecite{olivieri_vibrations_2017}). 
This suggests that the common-mode suppression in this frequency range is effective. 
In contrast, a few peaks are observed between $100\,\mathrm{Hz}$ and $1\,\mathrm{kHz}$, attributed to the mechanical beam modes of the pillars between each temperature plate or resonance modes of the active damping system. 
Then starting from slightly less than $1\,\mathrm{kHz}$, the spectra follow the $\mathcal{O} (f^{-2})$ dependence, as reported. \cite{olivieri_vibrations_2017} 
Eventually, they reach the detection limit of our interferometer setup at approximately $7\, \mathrm{pm} \mathrm{Hz}^{-1/2}$, which turned out to be limited by the noise floor of the oscilloscope, $\sim 6\,\mathrm{mV}_{\mathrm{rms}}$. 
It would have been possible to reach an order of magnitude better sensitivity, where the limit would be set by the noise equivalent power of the photodiode (PD). 

\section{\label{sec:Results_locking}Results of Cavity Stabilization}

\begin{figure*}[htbp]
    \includegraphics[width=\hsize]{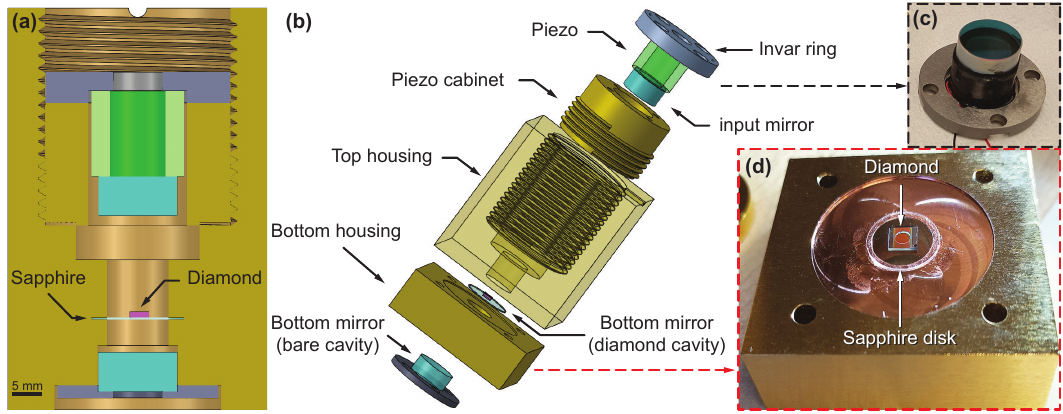}
    \caption{Optical cavity device. 
    (a) Cross-sectional view and (b) CAD rendering of the optical cavity housed in a gold-plated copper enclosure. 
    The input mirror, an AR-HR coated mirror on a piezo actuator, is mounted on an invar ring [gray, photograph in (c)] and fixed to a threaded copper support for cavity length adjustment. 
    The bottom mirror used for the `bare-cavity' demonstration (Subsection \ref{subsec:bareCavLocking}) is also mounted on an invar ring and fixed to the bottom cavity housing. 
    For the `diamond-integrated cavity' demonstration (Subsection \ref{subsec:DiaCavLock}), the AR-HR coated diamond itself acts as the bottom mirror, mounted on a sapphire disk for thermalization and alignment using a small amount of vacuum grease [photograph in (d)]. 
    Note that the standard bottom mirror is removed, and the diamond crystal is inserted only in the diamond-integrated configuration. Conversely, in the bare-cavity configuration, the diamond is completely removed from the optical path.}
    
    \label{fig:3_optical cavity design} 
\end{figure*}

\subsection{Optical Cavity Device and Locking Setup}\label{subsec:device}

Figure \ref{fig:3_optical cavity design} shows the design of the optical cavity device, which consists of two highly reflective mirrors. 
The optical cavity was configured in two variants. In the bare cavity setup, a standard flat HR mirror ($R\approx99\,\%$) was used as the bottom mirror. In the diamond-integrated configuration, this HR mirror was removed and replaced with a bulk diamond crystal with AR and HR coatings, mounted on a sapphire disk. The diamond crystal was physically removed from the cavity when operating in the bare cavity mode, ensuring that no additional loss or scattering from the diamond affected those measurements. 
The top mirror is a concave, highly reflective (HR) mirror with a reflectivity of $R\approx99\,\%$ and radius of curvature $250\,\mathrm{mm}$ used in both configurations to form a concave--flat cavity structure. Consequently, the beam waist is formed on the flat bottom mirror.~\mbox{\cite{kogelnik_laser_1966}}
The top mirror is attached to a piezo actuator (Piezomechanik, HPSt 150/14-10/12 with Black Coating). The piezo actuator enables \textit{in-situ} cavity length tuning via applied voltage and is mounted on an invar ring to minimize mechanical contraction upon cooling. 

The nominal distance between the two mirrors is designed to be approximately $30\,\mathrm{mm}$, resulting in a free spectral range (FSR) of about $5\,\mathrm{GHz}$. 
The mirror spacing can be globally adjusted using a cabinet with screw threads on its outer wall, which fits inside the enclosure housing, as shown in Fig. \ref{fig:3_optical cavity design} (b). 
This design enables tuning of the FSR from approximately $4.6\,\mathrm{GHz}$ to $6.5\,\mathrm{GHz}$ for the bare cavity and from $5.5\,\mathrm{GHz}$ to $9\,\mathrm{GHz}$ for the diamond-integrated cavity.

The optical parameters listed in 
Table \ref{tab:cavityparameters}
, particularly the cavity finesse and length, were chosen based on the anticipated requirements for future spin ensemble-based quantum transducers. 
In such applications, the cavity length sets the free spectral range (FSR)~\mbox{\cite{kogelnik_laser_1966}}, which should match the microwave transition frequency ($4 - 8\mathrm{~GHz}$) to enable resonant interaction and efficient photon conversion. 
A higher finesse~\mbox{\cite{suter_calculation_2014}} narrows the cavity linewidth, thereby increasing cooperativity~\mbox{\cite{pallmann_highly_2023}}, but at the cost of reducing the conversion bandwidth. 
Therefore, a moderate finesse on the order of a few hundred maintains a reasonable balance between sufficient cooperativity and operational transduction bandwidth. 
A detailed discussion of the limitations on the cavity finesse is provided in Subsection~\mbox{\ref{subsec:Loss}}.

\begin{table}[htbp]
\caption{
    Summary of cavity parameters. 
    }
    \centering
    \def\arraystretch{1.5}
    \begin{tabular}{ccc}
    \hline\hline
    \parbox[c]{3.4cm}{\centering Parameter} & \parbox[c]{2.4cm}{\centering Bare} & \parbox[c]{2.4cm}{\centering Diamond-integrated} \\
    \hline
    \parbox[c]{3.4cm}{\centering Cavity length (mm)}  & 32.5 & 27.3\\
    \parbox[c]{3.4cm}{\centering FSR (GHz)}     & 4.6 & 5.5   \\
    \parbox[c]{3.4cm}{\centering Finesse, $\mathscr{F}$}      & 310   & 90 \\
    \parbox[c]{3.4cm}{\centering Round trip loss (\%)} & 2.00 & 6.75\\
    \parbox[c]{3.4cm}{\centering Linewidth (MHz)} & 15 & 60\\
    \parbox[c]{3.4cm}{\centering Quality factor, $Q$}  & $2.7\times 10^7$ & $6.5\times 10^6$\\
    \parbox[c]{3.4cm}{\centering Beam waist in radius ($\mathrm{\mu m}$)}         &140    & 135\\
    \parbox[c]{3.4cm}{\centering Mode Volume ($\mathrm{mm^3} $)} &0.50 &0.39\\

    \hline\hline
    \end{tabular}
\label{tab:cavityparameters}

\end{table}

\begin{figure}[htbp]
    \includegraphics[width=\hsize]{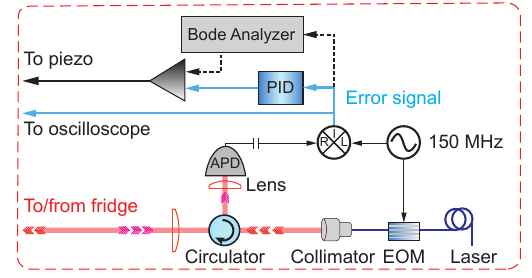}
    \caption{
    PDH locking setup for the cavity. 
    A frequency-stabilized laser is phase modulated by an electro-optic modulator (EOM) and is directed into the fridge. Reflected light from the cavity is separated by an optical circulator and detected by an avalanche photodiode (APD). The detected signal generates an error signal through a mixer, fed into a digital PID controller, which provides feedback to the piezo actuator. The frequency response of the locked cavity is measured using a RedPitaya in Bode analyzer mode. All the optics are aligned on the fridge-top optical breadboard to maintain the vibrations in the common-mode.
    }
    \label{fig:4_PDH locking scheme} 
\end{figure}

We employed the Pound-Drever-Hall (PDH) method \cite{black_introduction_2000} to lock the resonance frequency of the cavity to the laser frequency, as depicted in Fig. \ref{fig:4_PDH locking scheme}. 
The `locking' laser beam, phase-modulated at $150\,\mathrm{MHz}$ by a fiber electro-optic modulator (EOM), enters the cavity, and the reflected beam out of the fridge is routed through an optical circulator and detected by an avalanche photodiode (APD). 
The resulting electrical signals are mixed with the modulation signal to produce error signals, which are then processed by a digital PID circuit to generate a feedback signal. 
This feedback signal is amplified and directed back to the piezo actuator at the MXC stage to counteract the cavity displacement. 
Details of the setup are provided in the Appendix \ref{appendix:subsec:CavityLockingSetup}. 
All the cavity locking results presented in the following sections are measured at PT on and AD on condition unless mentioned otherwise. 

\subsection{\label{subsec:bareCavLocking}Locking a Bare Cavity to a Laser}

Using the setup, we stabilized a bare optical cavity—--without the inclusion of diamond--—at millikelvin temperature. 
To achieve stable locking of the cavity with a finesse $\mathscr{F}$, the root mean square (rms) of the cavity length fluctuation, $\delta L_{\mathrm{rms}}$ must be suppressed to:~\cite{vadia_open-cavity_2021}
\begin{equation} \label{eq:FvsLengthFluctuation}
    \delta L_{\mathrm{rms}} \leq \frac{\lambda}{2\mathscr{F}}, 
\end{equation}
where $\lambda$ is the wavelength of the laser \cite{}.

To identify the optical cavity spectrum, we initially scanned the cavity length by applying a triangular voltage to the piezo actuator. 
The transmitted signal amplitude and the error signal are represented by the dashed blue and orange curves, respectively, on the right side of the top panel in Fig. \ref{fig:5_BareCavityFluctuation}(a).
The two smaller sideband peaks in the blue curve originate from the phase modulation. 
From this and a wider scan (Fig. \ref{figS:11_CavityScanAnalysis} in Appendix \ref{appendix:subsec:analysis_cavLengthFluctuation}), the finesse and FSR are estimated to be approximately 310 and \textbf{$4.6\, \mathrm{GHz}$}, respectively. 

The blue data plotted in Fig. \ref{fig:5_BareCavityFluctuation}(a) shows the transmission signal when the cavity is locked, maintaining near-maximum amplitude over several seconds. 
We further continued, and the lock was sustained for over an hour (Fig. \ref{figS:13_LongCavityLocking} in Appendix \ref{appendix:subsec:analysis_cavLengthFluctuation}). 
The solid orange curve represents the error signal, which was converted to cavity length displacement and plotted in the bottom panel. 
The root mean square (rms) of the average cavity length fluctuation was determined to be approximately $27.6\pm 3\,\mathrm{pm}$, indicating that the setup can sustain an optical cavity with a finesse of approximately $1.3\times10^4$ at a wavelength of $\lambda = 737\, \mathrm{nm}$, as estimated from Eq.~\eqref{eq:FvsLengthFluctuation}. 
The details of the analysis are provided in the Appendix \ref{appendix:subsec:analysis_cavLengthFluctuation}. 

Note that the base temperature of our fridge, when the piezo actuator is shorted and the laser shutter is closed, is around $10\,\mathrm{mK}$. 
This base temperature was monitored at the mixing chamber (MXC) plate, located approximately 30 cm from the device. 
Although the thermometer was not attached to the device, it was rigidly anchored via an oxygen-free copper housing, ensuring good thermal contact~\mbox{\cite{valois_characterization_2024}} with the MXC. 
We therefore expect the device temperature to be close to that of the base plate.

We observed that applying a DC voltage of approximately 75 V to the piezo actuator increased the MXC temperature from 10 mK to 15 mK. To investigate whether this temperature rise was caused by high-frequency electrical noise from the piezo driver, we inserted a 2.5 kHz low-pass filter at room temperature (corresponding to about one-tenth of the piezo’s resonance frequency). Under these conditions, no measurable change in temperature was observed when applying the same DC bias, suggesting that the contribution from electrical noise is negligible and that the heating is primarily due to resistive loss in the piezo element itself.

We also evaluated the effect of laser-induced heating by varying the incident laser power and monitoring the resulting thermal load on the refrigerator. During the actual locking experiments, an optical power of approximately $10\mathrm{~\mu W}$ was used. This input laser power corresponds to a thermal load of approximately $0.4\mathrm{~\mu W}$, which is significantly smaller than the estimated $\approx6.3\mathrm{~\mu W}$ heat load from piezo actuator at 15 mK. No observable change in the base temperature was detected when the laser was turned on under these operating conditions.

\subsection{\label{subsec:DiaCavLock}Locking a Diamond-integrated Optical Cavity}

\begin{figure*}[t]
    \includegraphics[width=\textwidth]{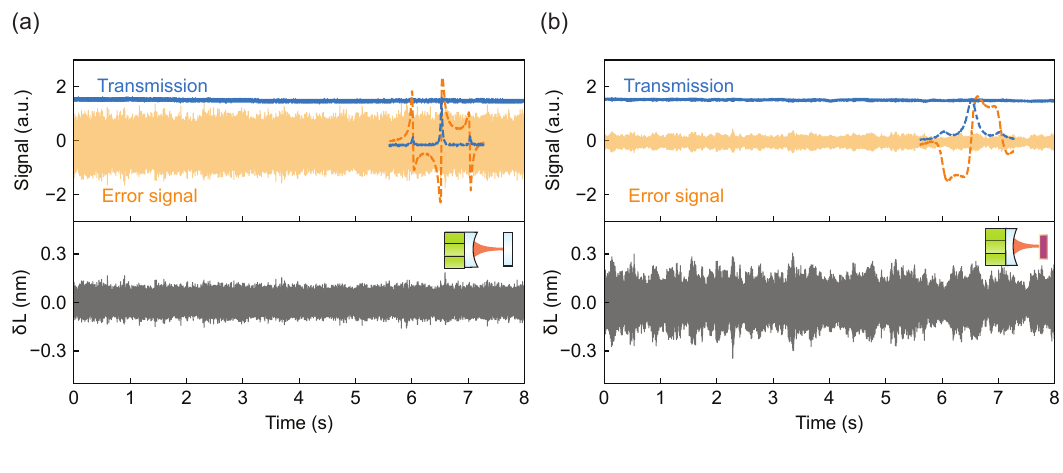}
    \caption{
    Cavity stability measurement at $15\,\mathrm{mK}$ for (a) the bare cavity and (b) the diamond-integrated cavity. 
    Upper panels: error signal (orange) and transmitted signal (blue) for a cavity length scan (dashed lines) and under lock (solid lines). 
    Lower panels: cavity length fluctuations derived from the error signals. The rms cavity length fluctuation is $\delta L_{\mathrm{rms}} \approx30~\mathrm{pm}$ for the bare cavity and $\approx63~\mathrm{pm}$ for the diamond-integrated cavity for this particular dataset.
    The higher finesse of the bare cavity corresponds to better sensitivity, resulting in larger error signal amplitudes in (a) compared to (b). 
    } 
    \label{fig:5_BareCavityFluctuation} 
\end{figure*}

Next, we investigated an optical cavity incorporating a bulk single crystal diamond (type-IIa) that is designed to develop a spin ensemble-based quantum transducer, as shown in Fig. \ref{fig:3_optical cavity design} (d). The crystal has dimensions of $4.5\times4.5\times0.5\, \mathrm{mm}$, with substitutional nitrogen impurities concentration of less than $1\,\mathrm{ppm}$ (The details of the sample are provided in Appendix \ref{appendix:subsec:Sample}).
However, bulk diamond crystals possess a high refractive index of $2.4$ at the wavelength of $737\,\mathrm{nm}$, which leads to significant Fresnel loss. 
To address this issue, previous studies have primarily focused on using thin diamond crystals, as thin as a few micrometers, \cite{janitz_fabry-perot_2015, raman_nair_amplification_2020} or positioning a diamond crystal at the Brewster angle. \cite{savvin_nv_2021} 

As an alternative to these approaches, we coated one side of the diamond with an anti-reflective (AR, $R <0.25\%$) coating to mitigate the reflection, \cite{dam_optimal_2018, xu_enhancement_2020, yeung_anti-reflection_2012} which appears better suited for the use of a bulk diamond crystal. 
The other side of the diamond was coated with a highly reflective (HR, $R \approx99\%$) coating, serving as one mirror of the Fabry-P\'erot cavity. 
Thanks to the AR coating, we did not observe `diamond-like' or `air-like' modes reported in cavities containing thin diamonds. \cite{janitz_fabry-perot_2015, raman_nair_amplification_2020,bogdanovic_design_2017} 
Nevertheless, the birefringence of the diamond \cite{luo_absorption_2023} caused a splitting of the transmitted signal, which we suppressed by adjusting the polarization using a half-wave plate mounted on top of the fridge. 

Following the protocol used for the bare cavity, we obtained the cavity resonance peaks and subsequently locked the diamond-integrated cavity. 
The transmission and error signals, while the piezo was scanned, are displayed on the top right of Fig. \ref{fig:5_BareCavityFluctuation}(b). 
While well-separated and distinct sideband peaks are not observed for the diamond-integrated cavity due to its internal losses, 
the finesse and FSR could still be measured to be approximately $90$ and $5.5\,\mathrm{GHz}$, respectively. 
The transmission and error signals under locking are indicated in Fig. \ref{fig:5_BareCavityFluctuation}(b) with solid blue and orange lines, respectively. 
The cavity length displacement during locking is plotted in the bottom panel of Fig. \ref{fig:5_BareCavityFluctuation} (b). The average rms value of the displacement, estimated over multiple measurements, is $\approx 69.3\pm 3.6\,\mathrm{pm}$, indicating that the setup can lock the diamond-integrated cavity with a finesse of approximately $5.3 \times 10^3$ at $\lambda = 737\,\mathrm{nm}$ from Eq.~\ref{eq:FvsLengthFluctuation}. 

\subsection{\label{subsec:Loss}Origin of the Internal Loss in the Diamond-Integrated Cavity}

The measured finesse of $\mathscr{F}\approx90$ corresponds to a total cavity loss of $\approx 6.8\,\%$ per round trip\cite{suter_calculation_2014} , which exceeds the mirror transmission loss of $2\,\%$ per round trip. 
To investigate the origin of the additional loss of $4.8\,\%$, we first considered surface scattering loss on the diamond. 
To this end, the roughness of the diamond surface was analyzed using atomic force microscopy before and after the optical coating, as explained in the Appendix \ref{appendix:subsec:Sample}. 
The root mean square roughness was measured to be $R_q = 1.5\,\mathrm{nm}$, enabling us to estimate the total integrated scattering loss ($S$) using the relation: \cite{bennett_recent_1992} 
\begin{equation}\label{eq:scattering loss}
    S \approx \left( \frac{4\pi R_q}{\lambda/n} \right)^2,
\end{equation}
where $n = 2.4$ is the refractive index of diamond. 
The loss per round trip due to the surface scattering is estimated to be $\approx 0.8\,\%$, leaving the majority of the total loss of approximately $4\,\%$ unexplained. 
Clipping loss was also evaluated and found to be negligible, as the diameter of the beam waist $2w_0 \approx 270\,\mu\mathrm{m}$ is much smaller than the aperture of the dielectric coating on the diamond substrate, $D = 3\,\mathrm{mm}$. This results in a clipping loss of\cite{hunger_fiber_2010} $e^{-2(D/2w_0)^2}\approx0$.
An off-centered beam due to misalignment could technically cause clipping loss. However, the finesse of $\approx 90$ is reproduced multiple times, even with small displacements of the diamond mirror each time. Therefore, we rule out the possibility of any clipping loss in our diamond-integrated cavity.

Based on the estimated scattering and clipping losses, we conclude that the internal loss of $4\,\%$ is most likely due to absorption inside the diamond, attributed to sub-bandgap defects.\cite{luo_absorption_2023} 
The absorption loss is given by $1-10^{-2d \alpha }$, where $\alpha$ and $d$ represent the absorption coefficient and the diamond thickness, respectively. 
From this, the absorption coefficient is estimated to be $\alpha \approx 0.15\,\mathrm{cm}^{-1}$, which matches well with the results measured at a laser wavelength of approximately $737\,\mathrm{nm}$ for low nitrogen concentration diamond samples reported in Ref.  \onlinecite{luo_absorption_2023}. 
We note that this absorption loss could be mitigated by using diamonds with lower defect densities,\cite{luo_absorption_2023} which would enable the realization of a higher finesse diamond-integrated cavity. 
Using a thinner diamond crystal would be another viable approach. 

Lastly, we observed that the finesses of both the bare and diamond-integrated cavities remained nearly constant from room temperature to low temperatures, suggesting that the optical coatings on the bulk mirrors and diamond were minimally affected by temperature changes.

\subsection{Cavity Length Fluctuation under Different Temperatures}
\label{subsec:Cavity_Temperature}

Cavity length fluctuations, $\delta L_{\mathrm{rms}}$, while locked may vary under different system temperatures. 
To assess this, we measured $\delta L\mathrm{_{rms}}$ for both cavities, the bare and diamond-integrated, at three different temperatures: room temperature (RT), $4\,\mathrm{K}$, and $15\,\mathrm{mK}$, with the PT on and off, while AD was always on. 
The results are summarized in Table \ref{tab:compare fluctuation}. 
The rms length fluctuation for the bare cavity remains approximately $\mathrm{30\,pm}$ ($\mathrm{20\,pm}$) during the PT on (PT off) conditions, as the device temperature varies from $4\,\mathrm{K}$ to millikelvin. 

In contrast, the diamond-integrated cavity exhibits slight variations in $\delta L_{\mathrm{rms}}$ when the temperature changes from $4\,\mathrm{K}$ to $15\,\mathrm{mK}$. 
Also, the corresponding $\delta L\mathrm{_{rms}}$ value is significantly higher for the diamond-integrated cavity than the bare cavity when the PT is on. 
Currently, we attribute the larger fluctuation to the relatively weak clamping of the sapphire disk to the bottom housing, which will also be discussed in the next Subsection. 
\begin{table}[htbp]
\caption{
    Comparison of root-mean-square (rms) cavity length fluctuations, $\delta L_{\mathrm{rms}}\,$ in picometers, for bare and diamond-integrated cavities at different temperatures. 
    PT stands for pulse tube cooler. 
    }
    \centering
    \def\arraystretch{1.5}
    \begin{tabular}{ccccc}
    \hline\hline
    \parbox[c]{2cm}{\centering $\delta L_\mathrm{rms}$ (pm)} & \multicolumn{2}{c} {Bare} & \multicolumn{2}{c} {Diamond-integrated}  \\
    \parbox[c]{2cm}{\centering Temperature}& \parbox[c]{1.4cm}{\centering PT on}  & \parbox[c]{1.4cm}{\centering PT off} &  \parbox[c]{1.4cm}{\centering PT on} & \parbox[c]{1.4cm}{\centering PT off}  \\
    \hline
    RT         &  -      & 23.5     &  -     & 13.1   \\
    4 K         & 31.6 $\pm 0.2$    & 19.8   &  46.0 $\pm 0.7$    & 21.5 \\
    15 mK         & 27.6 $\pm 3$      & 19.9   &  69.3 $\pm 3.6$    & 16.8 \\
    \hline\hline
    \end{tabular}
    \label{tab:compare fluctuation}
\end{table}

\begin{figure}[htbp]
    \includegraphics[width=\hsize]{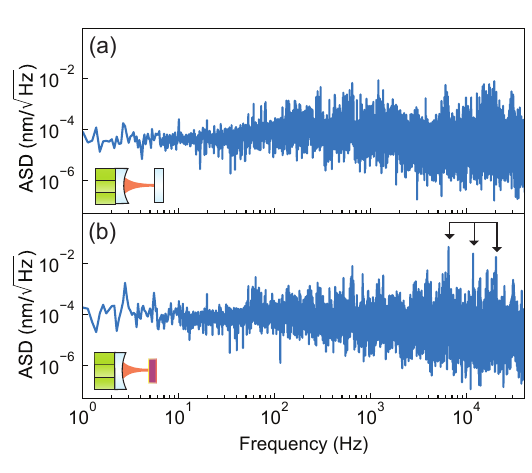}
    \caption{
    Cavity length fluctuations in frequency domain at $15\,\mathrm{mK}$. 
    (a) and (b) show the amplitude spectral density (ASD) for both the bare and diamond-integrated cavities, respectively. ASD is calculated from the fast Fourier transform (FFT) of the cavity length fluctuations, presented in the bottom panels of Fig. \ref{fig:5_BareCavityFluctuation}. Three prominent peaks were observed at approximately 6 kHz, 11 kHz, and 20 kHz in the diamond-integrated cavity, as indicated by arrows.
    }
\label{fig:6_frequencyresponse} 
\end{figure}

\begin{figure}[htbp]
    \includegraphics[width=\hsize]{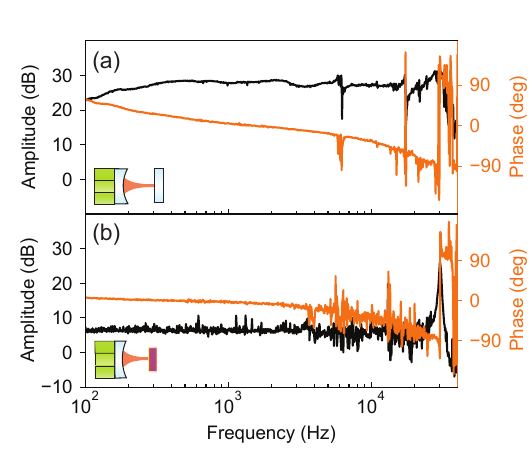}
    \caption{
    The frequency responses of cavity while locked at $15\,\mathrm{mK}$. 
    (a) and (b) display the Bode plots for the bare and diamond-integrated cavities, respectively. 
    The measurements were performed by sending a low-amplitude sinusoidal signal at each frequency to the piezo actuator. 
    }
\label{fig:7_Bodeplot} 
\end{figure}

\subsection{Mechanical Modes of the System}
\label{subsec:VibrationalModes}

The amplitude spectral density (ASD) of the relative vibrations, shown in Fig. \ref{fig:2_relative displacement}(c) in Subsection \ref{subsec:RelativeVibration}, provides insights into the mechanical vibrations at the device. However, it lacks sensitivity in the higher frequency range from around $1\,\mathrm{kHz}$, where the signal levels fall below the detection limit of our interferometer. 
In contrast, the length fluctuations $\delta L$ in the cavity-locked configuration, plotted in the bottom panels of Figs. \ref{fig:5_BareCavityFluctuation}(a) and (b), also contain information about such mechanical vibrations. 
To address this, we plot the ASD of the bare and diamond-integrated cavities in Figs. \ref{fig:6_frequencyresponse} (a) and (b). Because the error signal has higher sensitivity compared to the interferometer (Fig. \ref{figS:2_interferometerAnalysis} and Fig. \ref{figS:11_CavityScanAnalysis}), we can see the length fluctuation up to a few tens of $\mathrm{kHz}$. 

Several clusters of peaks emerge in the frequency range from $100\,\mathrm{Hz}$ to a few tens of $\mathrm{kHz}$, as previously reported. \cite{chijioke_vibration_2010,schmoranzer_cryogenic_2019,olivieri_vibrations_2017} 
We attribute these peaks to beam vibration modes of the fridge or resonance modes of the active damping system. 
Additionally, these peaks may be associated with the vibration modes of the optical cavity device and its supporting thermalization tube, as also observed in simulations (Appendix \ref{appendix:subsec:CavityMechanicalMode}).   

Although three distinct peaks were observed at approximately $6\,\mathrm{kHz}$, $11\,\mathrm{kHz}$, and $20\,\mathrm{kHz}$ in the diamond-integrated cavity [arrows in Fig. \ref{fig:6_frequencyresponse}(b)], only the $6\,\mathrm{kHz}$ peak has been reproduced across multiple cooldowns but it slightly moved around each time. 
This suggests that these modes are associated with the bottom-mirror-coated diamond, which is manually placed on a sapphire disk with a tiny amount of vacuum grease applied to the edge. 
This likely introduces changes in the cavity arrangement during each cooldown.

Finally, the piezo actuator may also contribute to the observed modes in the ASD plot. 
To investigate this, we measured the frequency response of the cavity for both the bare and diamond-integrated cavities under locked conditions using a Bode analyzer, as shown in Fig. \ref{fig:4_PDH locking scheme}. 
A low-amplitude sinusoidal signal in the frequency range from $100\,\mathrm{Hz}$ to $40\,\mathrm{kHz}$ was generated by the analyzer and applied to the piezo actuator. 
The error signal was then fed into the analyzer, enabling it to acquire the frequency response of the locked cavity. 
The results are shown in Figs. \ref{fig:7_Bodeplot} (a) and (b). 
Notably, unlike the ASD data, the Bode gain and phase data reveal several resonances only in the high-frequency region, especially around $6\,\mathrm{kHz}$, $18\,\mathrm{kHz}$, and $30\,\mathrm{kHz}$ for the bare cavity. 
For the diamond-integrated cavity, clusters of peaks centered around $6\,\mathrm{kHz}$ and $12\,\mathrm{kHz}$ were observed. 
Those peaks may arise from the mechanical resonances of the piezo actuator. 
Additionally, we note that the stycast adhesive between the piezo actuator and the invar ring may act as a spring, which could contribute to mechanical resonances in the $\sim \,\mathrm{kHz}$ range at cryogenic temperatures. \cite{kumar_quantum-enabled_2023}

\section{CONCLUSION}
We demonstrated the stabilization of two types of Fabry-P\'erot optical cavities, bare and diamond-integrated, at millikelvin temperatures in a custom cryogen-free dilution refrigerator. 
Mechanical vibrations were mitigated by placing an optical breadboard on top of the refrigerator, ensuring that the optical cavity device on the MXC plate and the optical setup vibrate in a common mode. 
The incorporation of a bulk diamond crystal was achieved by applying anti-reflective and highly reflective coatings. 
As a result, the cavity was locked to a free-running laser, a critical milestone towards practical quantum transducers. 
Average cavity length fluctuations of $27.6\pm 3\,\mathrm{pm}$ and $69.3\pm 3.6\,\mathrm{pm}$ were achieved for the bare and diamond-integrated cavities, respectively, corresponding to achievable finesses of $1.3\times 10^4$ and $5.3\times 10^3$. 
The primary source of loss in the diamond-integrated cavity was identified as absorption, which could potentially be mitigated by using a diamond crystal with fewer defects. 
This work introduces two key advances: stable cavity-to-laser locking at millikelvin temperatures, and successful stabilization of a cavity containing a bulk diamond crystal.
These results, achieved without major cryostat modifications, mark a significant step toward practical spin–based quantum transducers.

\begin{acknowledgments}

The authors thank P. Burns and J. Kindem for generously sharing their expertise on the optical cryostat setup, and R. Yamazaki and A. Noguchi for the valuable insights into cryo-optical cavities. 
The authors also thank S. Uetake for sharing his expertise in fabricating and assembling the invar components, as well as the electronics for cavity-lock. 
Additionally, we are grateful to the members of the Hybrid Quantum Device Team within the Science and Technology Group and the Experimental Quantum Information Physics Unit at Okinawa Institute of Science and Technology Graduate University (OIST) for their insightful discussions. 
The authors also acknowledge for the help and support provided by the Engineering Section of Core Facilities at OIST. 
This work has been supported by the JST Moonshot R\&D Program (Grant No. JPMJMS2066), JST-PRESTO (Grant No. JPMJPR15P7), Grant-in-Aid for JSPS Fellows (Grant No. 24KJ2178), The Nakajima Foundation, The Sumitomo Foundation, and OIST Graduate University.

\end{acknowledgments}

\section*{AUTHOR DECLARATIONS}
\subsection*{Conflict of Interest}
The authors have no conflicts to disclose.

\section*{Author Contributions}
T.H. and A.B. equally contributed to this work. 
Y.K. conceived the principle of the project. 
T.H. initiated the experiments. 
A.B., T.H. and Y.K. designed the device, conducted the experiments, and analyzed the data. 
Y.K. and H.T. jointly supervised the project. 
All authors contributed to the manuscript.

\section*{Data Availability Statement}
The data that support the findings of this study are available from the corresponding author upon reasonable request.

\clearpage
\appendix

\section{THE DILUTION REFRIGERATOR CONFIGURATION}
Here we provide detailed information about our dilution refrigerator setup that is designed to minimize the mechanical vibrations.
In addition, we introduce how to irradiate the laser from the top of the fridge to the device, which is anchored to the MXC stage. 
\hl{
Most of the vibration mitigation features described below are based on commercially available options from Bluefors, except for the optical breadboard (\mbox{~\ref{appendix:subsec:PassiveStabilization}}), the optical windows with narrow-band filters and anti-reflection coatings, and the custom threaded mounting tubes (\mbox{~\ref{appendix:subsec:OpticalPortSetup}}). 
}

\subsection{Passive Stabilization}
\label{appendix:subsec:PassiveStabilization}
We implemented passive mitigation strategies to address three primary mechanical noise sources anticipated in our cryogen-free dilution refrigerator system (Bluefors, LD400). 
The first one is acoustic noise from the helium compressor for the pulse tube cooler \hl{(Cryomech, PT415-RM)}, which can travel through the ground or air to the cryostat. 
To mitigate this, we placed the helium compressor in a separate `pump room', approximately 4 meters away from the fridge, and installed a soundproof wall between the rooms to reduce the acoustic noise. 
The second source is mechanical vibration noise transmitted to the cryostat via metal vacuum pipes connected to the turbo pump used for helium mixture circulation. 
We suppressed this noise by installing `T-dampers' at the top of the fridge and at the top of the turbo pump in the pump room. 
The final and seemingly largest source is the rotational motor valve of the pulse tube cooler. 
To mitigate this, the motor valve and buffer tanks are mechanically fixed and anchored to the building frame above the ceiling. 
Those countermeasures are illustrated in the photographs in Fig. \ref{figS:1_FridgeConstruction}. 

The optical breadboard, a custom-made product based on FB-162126(Y) by HERZ Co., LTD, is fixed on top of a $\mathrm{70\, mm}$-thick aluminum plate. 
The main body of our dilution refrigerator is also supported by this aluminum plate, facilitating the common-mode vibration rejection. 
To accommodate these heavy components, whose total weight exceeds the maximum load capacity allowed for the frame used for a standard LD-400, we used a thicker aluminum frame designed for larger systems, such as the XLD series by Bluefors. 

\subsection{Active Stabilization}
\label{appendix:subsec:ActiveStabilization}
\begin{figure}[ht]
    \includegraphics[width=\hsize]{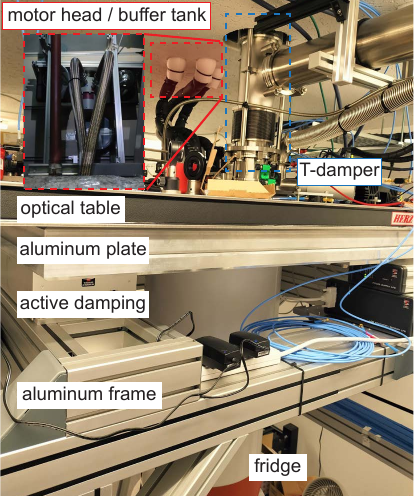}
    \caption{
    The custom dilution refrigerator and implemented countermeasures for vibration mitigation. 
    The photographs highlight key components, including the active damping system, T-dampers, and mechanical anchoring of the pulse tube cooler (red box). 
    See text for further details. 
    }
    \label{figS:1_FridgeConstruction} 
\end{figure}

In addition to the passive countermeasures, we implemented `active' mitigation to address vibrations from the ground, building structure, pedestrian traffic, and the rotational motor valve of the pulse tube cooler. 
An active damping system (Table Stable, AVI-400-M) was installed between the aluminum plate and the aluminum frame, as shown in Fig. \ref{figS:1_FridgeConstruction}. 
This system detects displacements and actively cancels them, within a frequency range of $1\,\mathrm{Hz}$ to $200\,\mathrm{Hz}$, as demonstrated in Fig. \ref{fig:1_Fridgesetup}(b) in the main text.

\subsection{Optical Port Setup}
\label{appendix:subsec:OpticalPortSetup}

\begin{figure}[ht]
    \includegraphics[width=\hsize]{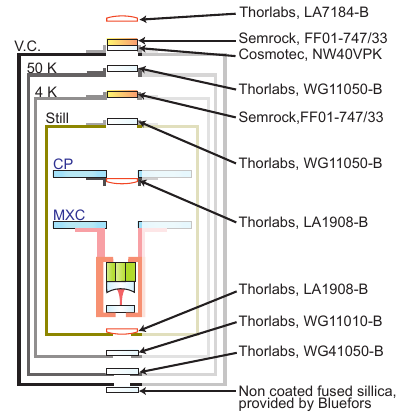}
    \caption{
    The combination of the optics inside dilution refrigerator 
    }
    \label{figS:2_FridgeOptics} 
\end{figure}

To enable optical access into the dilution refrigerator, we installed optical windows on each temperature stage and at the bottom of their radiation shields, as shown in Fig. \ref{figS:2_FridgeOptics}. 
Thermal radiation through the optical windows was minimized by reducing their diameter to less than 1.0 inch. 
We selected fused silica and BK-7 as the materials for the optical windows, as each blocks specific ranges of infrared light. To further reduce thermal radiation entering the fridge, narrow band-pass filter-coated windows were installed at the room temperature stage (vacuum-tight) and the 4 K stage. 
Anti-reflective (AR) coatings were applied to the windows at the other temperature stages to minimize reflective losses. 
All optical windows were mounted into custom-made `KF40 to SM1-threaded lens tube' flange adapters made of SUS304 stainless steel, which were fixed to the central line-of-sight flanges on respective temperature plates. 
The optical windows on the radiation shields, located at the bottom in Fig. \ref{figS:2_FridgeOptics}, were fixed by thin elastic metal plates. 

Figure \ref{figS:2_FridgeOptics} illustrates the combination of optics for each stage. 
Due to the long distance between fridge-top optical table and the device, we installed two lenses (f=500 mm), one at room temperature and the other at the cold plate inside the fridge, to construct the 4f-relay system.

\section{SAMPLE AND EXPERIMENTAL DETAILS}

\subsection{Samples}
\label{appendix:subsec:Sample}
The diamond sample used in this experiment is a single crystal synthesized by the chemical vapor deposition (CVD) method sourced from Element Six Technologies. 
The crystal has dimensions of 4.5 × 4.5 × 0.5 mm, with its surface aligned to the [001] crystallographic orientation.
Substitutional nitrogen impurities, known as P1 centers, are present at a concentration of less than $1\, \mathrm{ppm}$. 
The presence of any other color centers, such as nitrogen-vacancy centers, is anticipated to be negligible. 

The crystal underwent mechanical polishing (Syntek co., Ltd.) to remove any graphite layers, followed by cleaning in a boiling mixed-acid solution of $\mathrm{HNO_3}$, $\mathrm{H_2SO_4}$, and $\mathrm{HClO_4}$ in a 1:1:1 volume ratio for 10 minutes at 200 \textdegree C. \cite{brown_cleaning_2019} 
The acid treatment was performed within a fume hood, utilizing a setup similar to the one detailed in Ref. \onlinecite{brown_cleaning_2019}.
One face of the diamond was coated with an anti-reflective coating ($R<0.25\% @ 737\,\mathrm{nm}$), while the opposite face was coated with a highly reflective coating ($R=99\pm0.5\% @ 737\,\mathrm{nm}$). 
Both optical coatings were applied using ion beam sputtering (IBS) at Sigmakoki Co., Ltd. 

We evaluated the surface roughness of the diamond sample using an atomic force microscope (Bruker, Icon 3). As a measure of roughness, we estimate the root mean square deviation ($R_q$)  after each sample treatment and the corresponding values are $R_q$ $\approx 0.3 \,\mathrm{nm}$ after mechanical polishing and $R_q$ $\approx 1.5 \,\mathrm{nm}$ after optical coating using IBS.

\subsection{Cryogenic Piezo for Optical Cavity}
\label{appendix:subsec:cavitystructure}

The optical cavity, as elaborated in the main text, consists of two HR-coated mirrors, and the entire system is installed inside a gold-plated copper enclosure. We used Stycast (LOCTITE, 2850FTJ and CAT 23LVJ), an epoxy resin commonly used for cryogenic experiments, to fix the piezo actuator (Piezomechanik,HPSt 150/14-10/12 with Black Coating) and mirror to the invar ring plate. 
However, standard piezo actuators often cracked during the cooling cycles due to the mismatch in the coefficients of thermal expansion. 
This issue was resolved by using a piezo actuator (shown in Fig. \ref{fig:3_optical cavity design} (c) in the main text), which is coated with a thin layer of stycast across its surface. 
The coating effectively mitigates stresses due to thermal contraction during the cooling. 
The maximum stroke of the piezo actuators, with and without stycast coating, is approximately $2\,\mathrm{\mu m}$ at $15\, \mathrm{mK}$. This value is eight times shorter than at room temperature.

\subsection{Interferometer Experiment Setup}
\label{appendix:subsec:InterferometerSetup}
\begin{figure}[ht]
    \includegraphics[width=\hsize]{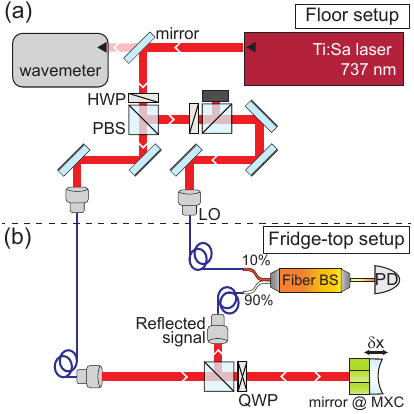}
    \caption{
    Interferometer setup for relative vibration characterization. 
    (a) Optical system on the ground-based optical table. 
    A small portion of the laser beam is directed to a wavemeter for wavelength monitoring, while the remaining is split into two using tunable beamsplitters, consisting of a half-waveplate (HWP; CASIX, WPZ1417-L/2-737) and a polarization beamsplitter (PBS), and then coupled into polarization-maintaining optical fibers. 
    (b) Optical system on the fridge-top optical breadboard. 
    The collimated laser beam from one of the optical fibers is directed to the cavity input mirror inside the dilution refrigerator. 
    The reflected laser beam is circulated by the combination of a quarter-waveplate (QWP; CASIX, WPZ1417-L/4-737) and PBS. 
    This reflected signal and the local oscillator signal interfere at the fiber beamsplitter, and the resulting signal is detected by a photodiode (PD). 
    }
    \label{figS:x_InterferometerSetup} 
\end{figure}

Figure \ref{figS:x_InterferometerSetup} illustrates the interferometer setup for relative vibration measurements. 
A Ti:Sapphire laser (M-squared, SolsTiS), placed on a ground-based optical table referred to as the `Floor setup' in Fig. \ref{figS:x_InterferometerSetup}, was tuned to approximately $737\,\mathrm{nm}$ and stabilized by locking it to an internal reference cavity. 
The stabilized laser beam was split into two, which were subsequently coupled into polarization-maintaining optical fibers [Fig. \ref{figS:x_InterferometerSetup}(a)]. 
On the fridge-top optical table, one of the collimated laser beams was directed to the cavity input mirror  mounted on the MXC stage inside the dilution refrigerator [Fig. \ref{figS:x_InterferometerSetup}(b)]. 
The reflected signal was routed through an optical circulator and combined with the local oscillator (LO) laser beam at a fiber beam splitter [(`Fiber BS' in Fig. \ref{figS:x_InterferometerSetup} (b);Thorlabs, PN850R2A1]. 
The resulting interference was then detected by a photodiode (PD; Thorlabs, PDA36A-EC). 
The absolute laser wavelength was continuously monitored by a wavemeter (HighFinesse, WS7-60). 

\subsection{Cavity Locking Experiment Setup}
\label{appendix:subsec:CavityLockingSetup}
We employed the standard PDH locking scheme to stabilize the cavity inside the dilution refrigerator. 
The experimental setup installed on top of the fridge is schematically illustrated in Fig. \ref{fig:4_PDH locking scheme} of the main text. 
A frequency-stabilized laser (M-squared, SolsTiS)) was phase-modulated at $150\,\mathrm{MHz}$ using a fiber-based electro-optic modulator (EOM; iXBlue, NIR-MPX800-LN-0.1). 
The modulated light was directed into the fridge via a series of mirrors and focused onto the optical cavity using a pair of lenses, each with a focal length of $500\,\mathrm{mm}$, arranged in a 4f configuration. 
The light reflected from the cavity was detected using a high-speed avalanche photodiode (APD; Thorlabs, APD430A/M).
The signal was passed through a bias tee (Mini-Circuits, ZFBT-4R2G-FT+) to separate the AC and DC components. 
The DC component was sent directly to an oscilloscope to monitor the reflected signal, while the AC component was filtered by a bandpass filter (Mini-Circuits, SBP-150+) and mixed with a reference signal from a signal generator using a frequency mixer (Mini-Circuits, ZLW-1-1+), producing the error signal. 
The error signal was then fed into a digital PID controller (TEM, LaseLock), which generates a feedback signal. 
This feedback signal was amplified by a bipolar piezo driver (MESS-TEK Co.,Ltd., M-2629) and sent back to the piezo actuator to lock the cavity. 
Additionally, the frequency response of the piezo actuator during locking was measured using an FPGA-based RedPitaya (STEMlab 125-14) in the Bode analyzer module. 
The dotted lines in Fig. \ref{fig:4_PDH locking scheme} indicate the connections to the RedPitaya. 

\section{DATA ANALYSIS}
In this section, we detail the analysis process for the relative vibration and the cavity length fluctuation while locking the cavity.

\subsection{Relative Vibration}
\label{appendix:subsec:analysis_relVibration}
Here, we detail the analysis and fitting methods used to determine the relative length fluctuations. 

\begin{figure}[ht]
    \includegraphics[width=\hsize]{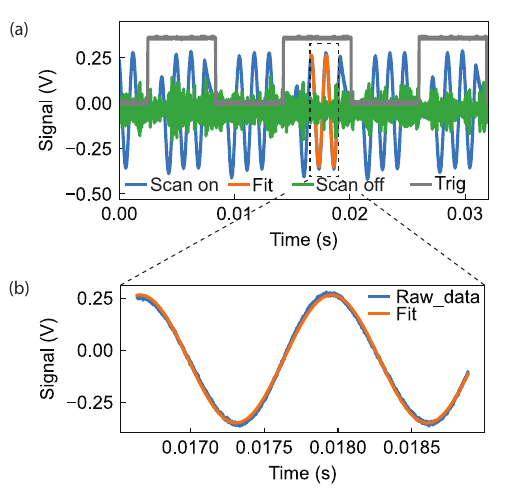}
    \caption{
    Analysis of relative vibration results. (a) Time-dependent scan on and scan off (relative vibration) signals are shown in solid blue and green colors, respectively}. A portion of the scan data is fitted with a sine function to determine the relative length fluctuation of the fridge. (b) A zoomed-in view of the sinusoidal fitting function is shown.
    \label{figS:2_interferometerAnalysis}
\end{figure}

The results in Fig. \ref{fig:2_relative displacement}(b) were analyzed as follows: 
to estimate the relative displacement in units of length, we scanned the mirror by applying a periodic triangular voltage to the piezo actuator. 
The resulting reflection data were fitted with a sinusoidal function described by:\cite{} 
\begin{equation}
    \label{eq:sin function}
    y = A\sin\left(2\pi ft+\phi\right)+D,
\end{equation}
where, $A$ is the amplitude, $f$ the interfering oscillation frequency, $\phi$ is the phase, and $D$ is the offset. 
The relative vibration signal (scan off), measured with the piezo actuator at zero applied voltage, was then converted into a length displacement ($\delta x$) using the following expression: 

\begin{equation}\label{eq:relVibrationCalibration}
    \delta x =\arcsin\left( \frac{y-D}{A}\right) \frac{\lambda}{4\pi},
\end{equation} 
where $\mathrm{\lambda}$ is the wavelength of the excitation laser. 
In Fig. \ref{figS:2_interferometerAnalysis}(a), the scan and relative vibration (scan off) data are shown as solid blue and solid green curves, respectively, while the solid orange line represents the sinusoidal fitting using Eq. \ref{eq:sin function}. 
A zoomed-in view of the fitting function is shown in Fig. \ref{figS:2_interferometerAnalysis}(b) for clarity.

Specifically, we note that our method requires the amplitude of the interferometric signal measured with no piezo actuation remains below the amplitude of the scanning signal used for calibration. This condition is satisfied in our measurements – the ``scan off'' signal (green trace in Fig.~\mbox{\ref{figS:2_interferometerAnalysis}}) is much smaller in amplitude than the scan on signal (blue trace in Fig.~\mbox{\ref{figS:2_interferometerAnalysis}}) used to fit the sine reference. Therefore, the conversion from voltage to displacement (Eq.~\mbox{\ref{eq:relVibrationCalibration}}) is valid for our data.

\begin{figure}[ht]
    \includegraphics[width=\hsize]{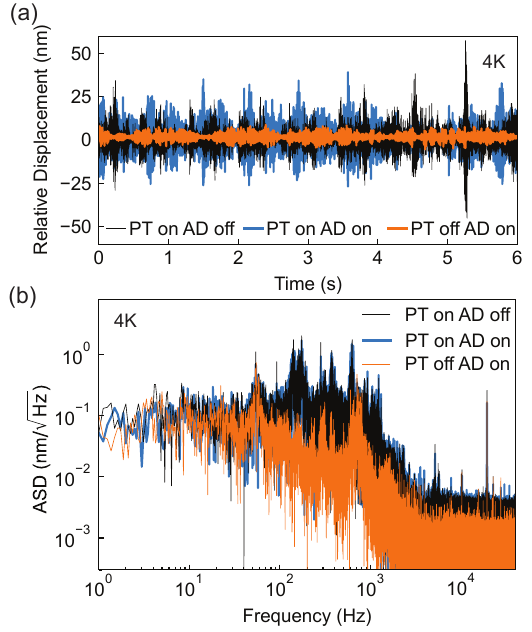}
    \caption{Relative vibration results measured at 4K. (a) Relative displacement measurements are shown for three different conditions: PT on and AD off, PT on and AD on, PT off and AD on. (b) Amplitude spectral density (ASD) measured from the displacement data show similar spectral behavior as measured at 15 mK temperatures, shown in the main text Fig. \ref{fig:2_relative displacement}(c).}
    \label{figS:RelVib4K} 
\end{figure}

As shown in Fig. \ref{fig:2_relative displacement}(b) in the main text, the peak-to-peak length fluctuations ($\delta L_{\mathrm{p-p}}$) are estimated to be $\mathrm{\approx 60\,\mathrm{nm}}$ when the PT is on, and $\mathrm{\approx 10\,\mathrm{nm}}$ when the PT is off. 
Moreover, we measured the relative displacement and the corresponding ASD spectral response estimated at 4K, as shown in Figs. \ref{figS:RelVib4K} (a) and (b), respectively. As mentioned in the main text, we hardly notice any significant difference between the ASD spectral response measured at 15 mK and 4K. Therefore, we infer that the turbo pump operating at $820 \,\mathrm{Hz}$, used for Helium circulation, has  minimal impacts on the displacement measurements.

\subsection{Cavity Length Fluctuation}
\label{appendix:subsec:analysis_cavLengthFluctuation}

\begin{figure}[ht]
    \includegraphics[width=\hsize]{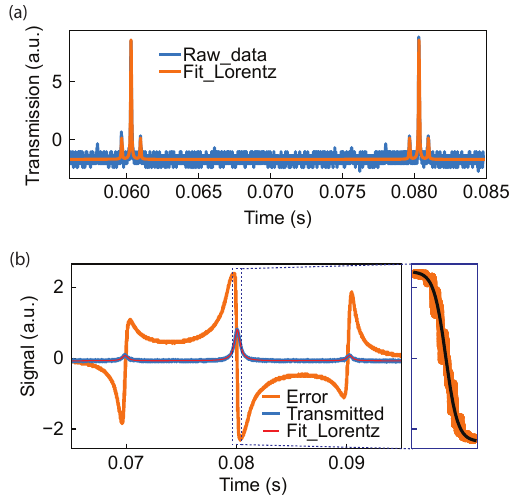}
    \caption{Analysis of the cavity scanning spectra. (a) A wide scan data produces the FSR value of the bare cavity in the transmission mode. The modes and the corresponding sidebands are fitted with three Lorentzian functions (solid orange line) to estimate the FSR and finesse value. (b) Transmitted and error signals are shown for the bare cavity. The slope of the error signal (encircled with a blue dashed line) is plotted separately and fitted with the hyperbolic tangent function (solid black line). The fitting enables to evaluate the conversion factor more precisely.} 
    \label{figS:11_CavityScanAnalysis} 
\end{figure}

Here, we detail the methods used to estimate cavity fluctuations in length. 
In Fig. \ref{figS:11_CavityScanAnalysis}(a), we present the cavity transmission obtained by scanning the bare cavity's length over a wide range to measure the free spectral range (FSR). 
The incoming laser is modulated at $150\,\mathrm{MHz}$ using an electro-optic-modulator (EOM) to generate sidebands. 
The $\mathrm{TEM_{00}}$ Gaussian modes and sidebands were fitted with sets of three Lorentzian functions each to determine the FSR value and the finesse ($\mathscr{F}$) of the cavity. 
The FSR (finesse) value for the bare cavity is approximately $4.6\,\mathrm{GHz}$ ($\mathscr{F}=310$), while for the diamond cavity, it is around $ 5.5\,\mathrm{GHz}$ ($\mathscr{F}=90$). 
In Fig. \ref{figS:11_CavityScanAnalysis}(b), the error signal and transmission for the bare cavity during scanning are shown. We fitted the transmission spectrum with Lorentzian functions (orange solid line).  
Additionally, the slope of the error signal, highlighted by the blue dashed box, is plotted separately on the right side of the main plot and fitted with a hyperbolic tangent function (black solid line). 
These fittings allow us to convert the error signal to the cavity length displacement using the cavity linewidth in length $(\Delta\mathrm{L})$, which follows the relation

\begin{equation}
    \label{eq:cavLengthCalibration}
    \Delta\mathrm{L} =\frac{\lambda} {2\mathscr{F}},
\end{equation}  
where $\lambda$ is the laser wavelength. 
With the finesse and the laser wavelength known, we converted the cavity linewidth from time to length unit. Finally, the error signal in voltage was converted into length fluctuations by analytically applying the inverse function of the hyperbolic tangent, as shown in the right panel of Fig. \ref{figS:11_CavityScanAnalysis}(b): 
\begin{equation}
    y=\frac{1}{2}\log \left(\frac{1+x}{1-x}\right),\, (-1<x<1)
\end{equation}

\begin{figure}[ht]
    \includegraphics[width=\hsize]{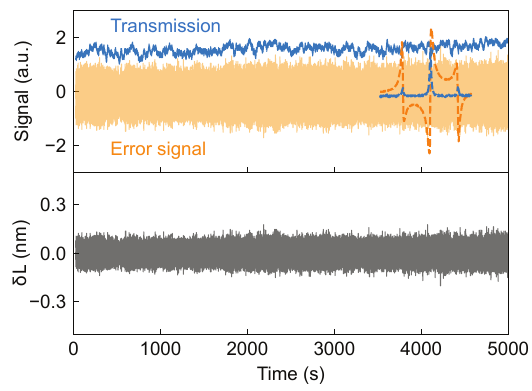}
    \caption{Stability of the bare cavity measured over an extended duration of time while locked at millikelvin temperature. 
    The upper panel shows the scanning (dashed lines) and locking (solid lines) data for the error (orange) and transmitted (blue) signals. 
    The bottom panel shows the cavity length fluctuation in unit of $\mathrm{nm}$ derived from the locked error signal.} 
    \label{figS:13_LongCavityLocking} 
\end{figure}

In the main article, we demonstrate cavity locking over a short duration. 
However, the cavity can remain locked even for hours without requiring re-locking. 
Here, in Fig. \ref{figS:13_LongCavityLocking}, the cavity stability is measured over a period exceeding one hour. 
The locked data remain stable throughout the measurement, with the observed slight deviation attributed to a drift in laser power. 
This result indicates the stability and robustness of the cavity locking. 

\subsection{Mechanical Modes of the Cavity}
\label{appendix:subsec:CavityMechanicalMode}

\begin{figure}[ht]
    \includegraphics[width=\hsize]{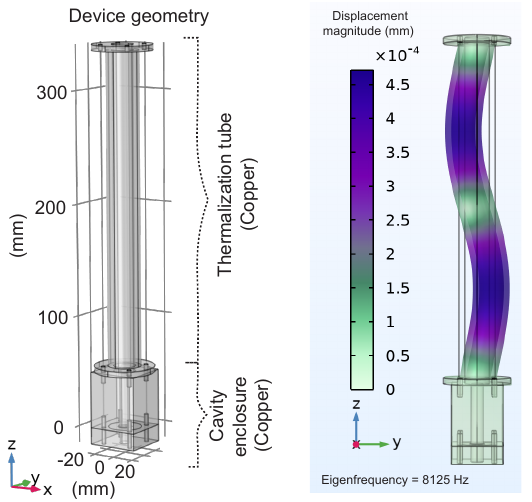}
    \caption{Mechanical modes of the device using COMSOL Multiphysics simulation. The schematic design in the left side displays the device geometry including the big copper thermalization tube. Right side image shows the mechanical mode of the whole system appearing around $8\,\mathrm{kHz}$}.
    \label{fig:comsolmode} 
\end{figure}

We performed numerical simulations to estimate the mechanical modes of the optical cavity device assembly using the Solid Mechanics module of COMSOL Multiphysics software. 
The simulation model replicated the real cavity structure, as shown in the left part of Fig. \ref{fig:comsolmode}. 
A long cylindrical copper thermalization tube (length of $\approx 280\,\mathrm{mm}$) was included, extending from the bottom of the MXC plate to position the sample at the center of the external magnetic field. 
We used the fixed constraint boundary condition across the physical joints to estimate the mechanical modes. 
The simulations were performed in the eigenfrequency modes, revealing mechanical modes ranging from a few $\mathrm{kHz}$ to tens of $\mathrm{kHz}$. 
An example of a simulated mechanical mode is displayed on the right-hand side of Fig. \ref{fig:comsolmode}, showing a mode around $8\, \mathrm{kHz}$. 
This mode is close to the intense peak around $6\,\mathrm{kHz}$, observed in both bare and diamond-integrated cavities, shown in Figs. \ref{fig:6_frequencyresponse}(a) and (b). 
However, the simulations could not reproduce the exact frequencies observed experimentally. 
This discrepancy is likely due to unknown parameters such as the spring constant of the vacuum grease at cryogenic temperatures, and the precise amount of grease used.

\bibliography{references}

\end{document}